\documentclass[a4paper]{article}
\usepackage{amsmath}

\newtheorem{theorem}{Theorem}

\newtheorem{example}{Example}
\newtheorem{definition}{Definition}

\begin{document}

\title{\textbf{Erratum to "Lattice constellation and codes from quadratic number fields" [IEEE Trans. Inform. Theory, vol. 47, No. 4, May. 2001}]}
\author{ Murat G\"{u}zeltepe  \\
%EndAName
{\small Department of Mathematics, Sakarya University, TR54187 Sakarya,
Turkey}}
\date{}
\maketitle

\begin{abstract}
We correct a partial mistake for a metric presented in the article
"Lattice constellation and codes from quadratic number fields"
[IEEE Trans. Inform. Theory, vol. 47, No. 4, May. 2001]. We show
that the metric defined in the article is not true, therefore,
this brings about to destroy the encoding and decoding procedures.
Also, we define a proper metric for some codes defined in the
article and show that there exist some $1-$error correcting
perfect codes with respect to this new metric.
\end{abstract}

%%% ----------------------------------------------------------------------



\section{Introduction and preliminaries}

In this Section, we show that the metric defined in \cite{Neto} is
not a true metric. Later, we define a proper Mannheim distance
over $A_p[w]$. Note that the matric given in  \cite{Neto} is
inspired by the Mannheim metric introduced in  \cite{Huber}.
Unfortunately, it is proved that the Mannheim metric is incorrect
in  \cite{Carmen}.

In \cite{Neto}, labeling procedure for the elements of $A_p[w]$ by
elements of the Galois field of order $p$, $GF(p)$, has been given
as follows:

i) Given a prime $p$ that splits completely over $\mathcal{Z}[w]$,
let $\pi=a+bw $ be a solution of $N(\pi)=\pi   \overline \pi =p$,
where $ \mathcal{Z}$ denotes the set of all integers, and
$\overline \pi$ denotes the conjugate of $\pi$.

ii) Let $s\in \mathcal{Z}$ be the only solution (in $r$) to the
equation $a+br\equiv 0,\ (mod \ p)$, where $0\le r \le p-1$.

iii) The element ${l}\in GF(p)$ is the label of the point $\alpha
= x+yw \in \mathcal{Z}[w]$ if $x+sy \equiv l\ (mod\ p)$ and
$N(\alpha)$ is minimum.

\begin{example} Let $d=-3$ and $p=7 \equiv 1 \ (mod \ 6)$.

i) A solution to the equation
$N(\alpha)=a^2+ab+\frac{1-d}{4}b^2=7$ is given by $(a,b)=(1,2)$.
Thus, we can take $\pi = 1+2w$.

ii)The only solution to the equation $1+2r\equiv 0 \ (mod\ 7)$,
where $0\le r \le 6$ is $3$.

iii) The element $l$ is the label of the point $\alpha = x+yw \in
\mathcal{Z}[w]$, if $x+3y\equiv l\ (mod \ 7)$ and $N(\alpha)$ is
minimum. Hence, the set $A_7[w]$ is obtained as $\left\{ {0, \pm
1, \pm w, \pm {w^2}=\pm \overline w} \right\}$. The set $A_7[w]$
is a finite field.

\end{example}

\begin{example} Let $d=-3$ and $p=193 \equiv 1 \ (mod \ 6)$.

i) A solution to the equation
$N(\alpha)=a^2+ab+\frac{1-d}{4}b^2=7$ is given by $(a,b)=(7,9)$.
Thus, we can take $\pi = 7+9w$.

ii)The only solution to the equation $7+9r\equiv 0 \ (mod\ 193)$,
where $0\le r \le 192$ is $85$.

iii) The element $l$ is the label of the point $\alpha = x+yw \in
\mathcal{Z}[w]$, if $x+85y\equiv l\ (mod \ 193)$ and $N(\alpha)$
is minimum. Some elements of the finite field $A_{193}[w]$ are
$9\equiv -7+7w$, $94 \equiv 2-8w$, $108\equiv -w$ $(mod \
(7+9w))$.

\end{example}

\begin{definition} \cite{Neto} Given an element $\gamma =x+yw \in A_p[w]$, the
Mannheim weight of $\gamma$ is defined as
$$W_M(\gamma)=\left| x \right|+\left| y \right|.$$ Also, the
Mannheim distance between any two elements $\alpha$ and $\beta$ in
$A_p[w]$ is defined as $$d_M(\alpha , \beta)=W_M(\delta),$$ where
$\delta \equiv \alpha - \beta\ (mod \ \left\langle \pi
\right\rangle),\ \delta \in A_p[w]$ with $N(\delta)$ minimum.

\end{definition}

But, $d_M(\alpha , \beta)=W_M(\delta)$ is not a true metric since
it does not fulfil the triangular inequality.

\begin{example} Let $d=-3$ and $p=193 \equiv 1 \ (mod \ 6)$. Then,
$\pi = 7+9w$ and $r=85$. Consider $A_{193}[w]$ and the elements
$x=-6+7w$, $y=1$, and $z=1-w$. The inequality
$$d_M(x,y)\le d_M(x,z)+d_M(z,y)$$ should be verified, but this is not true:

\begin{itemize}
    \item $d_M(x,y)=14$ since $x-y=-7+7w$ with minimum norm
    $N(-7+7w)=49$;
    \item $d_M(x,z)=10$ since $x-z=2-8w$ with minimum norm
    $N(2-8w)=52$;
    \item $d_M(z,y)=1$ since $z-y=-w$ with minimum norm
    $N(-w)=1$.
\end{itemize}

\end{example}

Now, we define a Mannheim metric over $A_p[w]$.

We denote the set of units in $A_p[w]$ by $\cal E$. It is easy to
check that $\mathcal E$ is the union of a set as indicated below:
\begin{equation}\label{eq:epsilon} \mathcal{E} = \left\{ { \mp 1, \mp \omega , \mp \overline \omega  }
\right\} .
\end{equation}

We note that for any two distinct elements $\epsilon_1$ and
$\epsilon_2$ in $\cal E$
\begin{equation}\label{eq:2}
N(\epsilon_1-\epsilon_2)\in\{1,2,3,4\}.
\end{equation}
Hence, if $\pi \overline \pi$ is equal to a prime number $p\geq
7$, $p \equiv 1\ (mod\ 6)$ we may conclude that the elements in
$\mathcal E$ represent $6$ distinct elements in $A_p[w]$.

Consider the direct product $S=A_p[w]^n$ of $n$ copies of
$A_p[w]$. We say that two elements, or \emph{words}, $\bar x$ and
$\bar y$ in $ A_p[w]^n$ have distance one, $\mathrm{d}_{m}(\bar
x,\bar y)=1$, if there is a word $\bar
e=(0,\ldots,0,\epsilon,0,\ldots,0)$, with just one non-zero entry
such that
\[
\bar y=\bar x+\bar e,
\]
for a unique element $\epsilon$ in a set $\mathcal{E}$.

With terminology from graph theory, it is now easy to explain how
we can define a metric in $A_p[w]^n$. Consider the words of $S$ as
vertices in a graph, where there is an edge between two vertices
$\bar x$ and $\bar y$ if $\mathrm{d}_{m}(\bar x,\bar y)=1$. The
\emph{distance} $\mathrm{d}_{m}(\bar a,\bar b)$ between any two
vertices $\bar a$ and $\bar b$ is the length of the shortest path
between these two vertices. General results from graph theory give
that this distance function defines a metric in $S$.

If $\mathcal E$ is defined as in Eq. (\ref{eq:epsilon}), then the
metric obtained in $A_p[w]^n$ is called the \emph{Mannheim
metric}.

We can give an alternative Mannheim metric which is equivalent to
above definition.

For this, we first give a modulo function from the Galois field
$GF(p)$ to the $A_p[w]$.

\begin{definition}  Let $\pi = a+bw$ such that $\pi \overline \pi =p=a^2+ab+b^2 \equiv 1 \ (mod \
(6))$, where $p$ is a prime and $a,b\in \mathcal{Z}$. We define
the modulo function $\mu :GF\left( p \right) \to {A_p}\left[ w
\right]$ as $$\mu \left( l \right)  = \left\{
{\begin{array}{*{20}{c}}
   {x + yw,}  \\
   {{x^{'}} + {y^{'}}\overline {w}, }  \\
\end{array}} \right.\quad \begin{array}{*{20}{c}}
   {\left| x \right| + \left| y \right| \le \left| {{x^{'}}} \right| + \left| {{y^{'}}} \right|}  \\
   {\left| x \right| + \left| y \right| > \left| {{x^{'}}} \right| + \left| {{y^{'}}} \right|}  \\
\end{array}.$$
Here, $x+ry\equiv l \ (mod \ p)$ and $x + yw={x^{'}} +
{y^{'}}\overline {w}$, where $a+br\equiv 0 \ (mod \ p)$, $0\le
r\le p-1$.
\end{definition} For example, $w^2=-1+w=-\overline {w}$. So, $x , y,{x^{'}}
,{y^{'}}$ are $-1,1,0,-1$, respectively.

\begin{example} Let $p=7 \equiv 1 \ (mod \ 6)$. Then, $\pi = 1+2w$.
The only solution to the equation $1+2r\equiv 0 \ (mod\ 7)$, where
$0\le r \le 6$ is $3$. Thus, we obtain the elements of $A_7[w]$
using by the modulo function $\mu$ as

$$\begin{array}{*{20}{c}}
   {\mu (0) =\ 0;}  \\
   {\mu (1) =\ 1;}  \\
   {\mu (2) =  - \overline w;}  \\
   {\mu (3) =\ w;}  \\
   {\mu (4) =  - w;}  \\
   {\mu (5) =\ \overline w;}  \\
   {\mu (6) =  - 1.}  \\
\end{array}$$
Hence, we obtain ${A_7}[w] = \left\{ {0, \pm 1, \pm w, \pm
\overline w } \right\}$.
\end{example}

\begin{definition} Given an element $\gamma=x+yw=x^{'}+y^{'}\overline
w$ in $A_p[w]$, we define the Mannheim weight of $\gamma$ as

$${W_m}(\gamma ) = \left\{ {\begin{array}{*{20}{c}}
{\left| x \right| + \left| y \right|,\,\left| x \right| + \left| y \right| \le \left| {{x^{'}}} \right| + \left| {{y^{'}}} \right|}\\
{\left| {{x^{'}}} \right| + \left| {{y^{'}}} \right|,\,\left| x
\right| + \left| y \right| > \left| {{x^{'}}} \right| + \left|
{{y^{'}}} \right|}
\end{array}} \right.$$

We also define the Mannheim distance between any two elements
$\alpha$ and $\beta$ in $A_p[w]$ as $$d_m(\alpha,
\beta)=W_m(\delta),$$ where $\delta \equiv \alpha - \beta \ (mod \
\pi)$, $\delta \in A_p[w]$.

\end{definition}

It should be noted that, in general, the Mannheim distance $d_M$
defined in \cite{Neto} and the Mannheim distance $d_m$ given here
are not isomorphic, as shown in the next example.

\begin{example}

Consider $A_p[w]^1$ and the elements 1 and $\pm w^2$ . We note
that

$$d_M(1,0)=W_M(1)=1=d_m(1,0)=W_m(1) $$

while
$$d_M(\pm w^2,0)=W_M(\pm w^2)=2 \ne 1=d_m(\pm w^2,0)=W_m(\pm w^2). $$

\end{example}

\section{$1-$Error-Correcting Perfect Codes}

In this section, $\beta $ will denote an element of order $6n=p-1$
such that $\beta ^n=w$. Thus, $\beta $ is a primitive element of
$A_p[w]$.

Let $p=6n+1$ be a prime in $\mathcal{Z}$ which factors in
$\mathcal{Z}[w]$ as $\pi \overline \pi$, where $\pi$ is a prime in
$\mathcal{Z}[w]$. Let $\beta$ denote an element of $${A_p}\left[ w
\right] \cong {{Z\left[ w \right]} \mathord{\left/
 {\vphantom {{\mathcal{Z}\left[ w \right]} {\left\langle \pi  \right\rangle }}} \right.
 \kern-\nulldelimiterspace} {\left\langle \pi  \right\rangle }}$$
 of order $6n$. Hence $\beta ^n =w$, and since $\beta$ is a primitive element
of $A_p[w]$, it can written  ${A_p}[w] = \left\langle \beta
\right\rangle  \cup \left\{ 0 \right\}$. Now let $C$ be the
null-space of the matrix

\begin{equation}  \label{eq:1}H = \left( {\begin{array}{*{20}{c}}
   1 & \beta  & {{\beta ^2}} &  \cdots  & {{\beta ^{n - 1}}}  \\
   1 & {{\beta ^7}} & {{\beta ^{14}}} &  \ldots  & {{\beta ^{7(n - 1)}}}  \\
    \vdots  &  \vdots  &  \vdots  &  \cdots  &  \vdots   \\
   1 & {{\beta ^{6t + 1}}} & {{{({\beta ^{6t + 1}})}^2}} &  \cdots  & {{{({\beta ^{6t + 1}})}^{(n - 1)}}}  \\
\end{array}} \right),\end{equation}
where $t<n$. An $n-$tuple  $$c = \left( {\begin{array}{*{20}{c}}
   {{c_0},} & {{c_1},} & { \cdots ,} & {{c_{n - 1}}}  \\
\end{array}} \right) \in A_p^n\left[ w \right]$$
is a codeword of $C$ if and only if $Hc^t=0$, where $c^t$ denotes
the transpose of $c$. If $ c(x) = \sum\nolimits_{i = 0}^{n - 1}
{c_i x^i }$ is the associated code polynomial, we get $$ c(\beta
^{6j + 1} ) = 0,{\rm for }j = 0,1, \cdots ,t.$$ The polynomial $
g(x) = (x - \beta )(x - \beta ^7 ) \cdots (x - \beta ^{6t + 1} )$
is the generator polynomial of $C$, and $ C = \left\langle {g(x)}
\right\rangle$ is an ideal of $ {{A_p[w][x]} \mathord{\left/
 {\vphantom {{H(K_1 )_\pi  [x]} {\left\langle {x^n -w} \right\rangle }}} \right.
 \kern-\nulldelimiterspace} {\left\langle {x^n  -w} \right\rangle
 }}$. If multiplying a code polynomial $c(x)$  by $x \ (mod (x^n  -w))$, we get $$xc(x) = c_0 x + c_1 x^2  +  \cdots  + c_{n - 1}
 x^n,$$ which belongs to $C$. We know that $x^n  =  w$. Therefore, if $c(x) \in
 C$, then $xc(x) \in C$. Thus, multiplying $c(x)$ by $x (\bmod (x^n -w))$ means the following:

\begin{enumerate}
    \item Shifting $c(x)$  cyclically one position to the right;
    \item Rotating the coefficient $c_{n-1}$ by $\pi/3$ radians in the complex plane
    and substituting it for the first symbol of the new codeword.
\end{enumerate}

Therefore, code $C$ defined by the parity check matrix in (1) is a
$w-$cyclic codes by considering a primitive root $\beta$ such that
$\beta ^n =w$.

\begin{theorem} Let $C$  be the null-space of the matrix \begin{equation}  \label{eq:1}H = \left( {\begin{array}{*{20}{c}}
   1 & \beta  & {{\beta ^7}} &  \cdots  & {{\beta ^{n - 1}}}  \\
\end{array}} \right).\end{equation}

Then  $C$ is able to correct any error pattern of the form $e(x) =
e_i x^i$, where $ W_{m} (e_i )=1$.
\end{theorem}

The proof of Thm. 1 is the same as the proof of Thm. 7 in
\cite{Neto}

Recall that the elements of Mannheim weight 1 of the alphabet
$A_p[w]$ are $\pm 1,\ \pm w, \ \pm \overline w$. By the
sphere-packing we get  $$p^{n-1}(6n+1)=p^{n-1}p=p^n.$$ Hence, the
codes defined by the parity check matrix in (4) are perfect.


\begin{thebibliography}{9}
\bibitem{Neto} T. P. da N. Neto, J. C. Interlando., "Lattice constellation and codes from quadratic number fields," IEEE Trans. Inform. Theory, vol. 47, No. 4, May. 2001.
\bibitem{Huber} K. Huber., "Codes Over Gaussian integers," IEEE Trans. Inform.Theory, vol. 40, pp. 207-216, Jan. 1994.
\bibitem{Carmen} C. Martinez, R. Beivide and E. Gabidulin., "Perfect codes for metrics induced by circulant graphs," IEEE Trans. Inform. Theory, vol. 53, No. 9, Sep. 2007.


\end{thebibliography}
\end{document}